\documentclass[12pt]{iopart}

\usepackage{iopams}
\usepackage{cite}
\usepackage{graphicx}
\usepackage{color}

\renewcommand{\vec}[1]{{\mbox{\boldmath$#1$}}}  
\begin{document}
\title[Relativistic calculations of charge transfer]{Relativistic calculations of charge transfer probabilities
in $ \mathbf{U^{92+} - U^{91+}(1s)}$ collisions using the basis set of cubic Hermite  
splines}
\author{I.A. Maltsev$^1$, G.B. Deyneka$^2$, I.I. Tupitsyn$^1$, V.M. Shabaev$^1$, Y.S. Kozhedub$^1$, G. Plunien$^3$, and Th. St\"{o}hlker$^{4,5,6}$}
\address{$^1$ Department of Physics, St. Petersburg State University, Ulianovskaya 1, Petrodvorets, 198504 St.Petersburg, Russia\\[0.2cm]}
 \address{$^2$ St. Petersburg State University of Information Technologies, Mechanics and Optics, Kronverk av. 49, 197101 St.Petersburg, Russia\\[0.2cm]}
  \address{$^3$ Institute f\"{u}r Theoretische Physik, Mommsenstrasse 13, D-01062 Dresden, Germany\\[0.2cm]}
 \address{$^4$ Gesellschaft f\"{u}r Schwerionenforschung, Planckstrasse 1, D-64291 Darmstadt, Germany\\[0.2cm]}
  \address{$^5$ Physikalisches Institut, Philosophenweg 12, D-69120 Heidelberg, Germany\\[0.2cm]}
  \address{$^6$ Helmholtz-Institut Jena, D-07743 Jena, Germany}
\ead{ilia.alexm@gmail.com}
\begin{abstract}
A new approach for solving the time-dependent two-center Dirac equation is presented. 
The method is based on using the finite basis set of cubic Hermite splines on a two-dimensional
lattice. The Dirac equation is treated in rotating reference frame. The collision of $\rm U^{92+}$ (as a projectile) 
and $\rm U^{91+}$ (as a target) is considered at energy $\rm E_{lab}=6$ MeV/u. The charge transfer probabilities 
are calculated for different values of the impact parameter. The obtained results are compared with 
the previous calculations [I.~I. Tupitsyn \textit{et al.}, Phys. Rev. A
\textbf{82}, 042701 (2010)], where a method based
on atomic-like Dirac-Sturm orbitals was employed. This work can provide a new 
tool for investigation of quantum electrodynamics  effects in heavy-ion collisions near the supercritical regime.
\end{abstract}
\section{Introduction}
Heavy-ion collisions can be used as a powerful tool for tests of
quantum electrodynamics at the supercritical Coulomb fields \cite{Greiner}.
To date many approaches were employed to treat such collisions~\cite{Becker86, 
Strayer90, Thiel92, Wells96, Pindzola_00, Busic_04, Momberger96, Ionescu99, 
Eichler90, Rumrich93, Momberger93, Gail2003, Tup_10}.
In our previous work \cite{Tup_10, Tup_12} a method of solving the time-dependent 
Dirac equation in the basis of atomic-like
Dirac-Fock-Sturm orbitals was developed. 

In the present paper we develop an alternative approach to treatment of 
low-energy heavy-ion collisions. The method is based on using the cubic Hermite spline 
basis set on a two-dimensional uniform grid. 
The Dirac equation is considered in a rotating 
reference frame. 
Previously the time-dependent Schr\"odinger equation was treated in a rotating coordinate
frame in Ref.~\cite{Grun82}. The basis of cubic Hermite splines was used for non-relativistic 
time-dependent calculations~\cite{Deineka_04, Deineka_06} and for relativistic time-dependent calculations in
the monopole approximation~\cite{Deineka_12}. To our best knowledge, however, the Hermite splines 
have not been employed before for solving the time-dependent Dirac equation for the real
two-center problem.   

The theoretical methods are described in section~\ref{sec:theory}. The results of our calculations of the
charge-transfer probabilities for the $\rm U^{92+}$--$\rm U^{91+}(1s)$ collision are presented in section~\ref{sec:Results}.
 Atomic units ($\hbar=e=m=1$) are used throughout the paper.
\section{Theory}
\label{sec:theory}
\subsection{Basis of the Hermite splines}
In this paper a basis of cubic Hermite splines is used. This basis can be defined in the following way.
The interval $[a,b]$ is partitioned  into $N$ equal subintervals $a= x_0 < x_1 < \ldots < x_N = b$,
with the length $h=x_{i+1}-x_i$ ($i=1,\ldots, N-1$). For each node $x_i$ two functions are introduced:
\begin{equation}
   s^0_i(x)=f^0 \left( \frac{x-x_i}{h} \right), \qquad s^1_i(x)=h f^1 \left( \frac{x-x_i}{h} \right)\,,
 \label{eq:h_basis}
\end{equation}
where
\begin{equation}		
    f^0(x)=\left\{
      \begin{array}{lc}
	0, \quad |x| \ge 1  \\
	(1-x)^2 (1+2x), \quad  0\le x<1\\
	(1+x)^2 (1-2x), \quad  -1<x<0\\
      \end{array}
      \right.
    \label{eq:spline0}
  \end{equation}

  \begin{equation}		
  f^1(x)=\left\{
    \begin{array}{lc}
      0, \quad |x| \ge 1  \\
      (1-x)^2 x,  \quad 0\le x<1\\
      (1+x)^2 x, \quad -1<x<0.\\
    \end{array}
    \right.\\
  \label{eq:spline1}
\end{equation}
Since only adjacent splines overlap, the Hamiltonian and overlapping matrices are sparse.
It is the significant advantage of this basis. The two-dimensional basis Hermite splines are defined as
\begin{equation}
 \phi_{j}(x,y) \equiv s^\mu_l(x)\cdot s^\nu_m(y).
  \label{eq:two_dim_splines}
\end{equation} 
Here $j=j(\mu,\nu,l,m)$ enumerates the splines $\phi_{j}(x,y)$ in the certain way. 

\subsection{Solution of the time-dependent two-center Dirac equation }
We consider a collision of $\rm U^{92+}$ (target) and $\mathrm{U^{91+}}(1s)$ (projectile). The position
of the target is fixed, the projectile moves along a straight line with a constant velocity.  

The electron movement is considered in the cylindrical coordinate system  $\left( \rho, z, \varphi \right)$ 
which is rotating around the target with the internuclear axis as $z$-axis.  
In this coordinate system the Dirac equation is written as~\cite{Muller76}  
\begin{equation}
i\frac{\partial \psi}{\partial t}=\hat{H}  \psi ,
\label{eq:rot_Dirac}
\end{equation}
\begin{equation}
 \hat{H}=\hat{H}_0-\vec{\hat{J}} \vec{\omega},
\end{equation}
where $\psi$ is the Dirac wave function, $\vec{\hat{J}}$ is the angular
momentum operator, 
$\vec{\omega}$ is the angular velocity of the reference frame, and
  
\begin{equation}
\hat{H}_0=c \vec{\alpha} \vec{\hat{p}}+\beta c^2+V(\vec{r},t).
\label{eq:inert_Hamiltonian}
\end{equation}
Here $\vec{\alpha}, \beta$ are the Dirac matrices, and $V(\vec{r},t)$ is the
two-center potential. In Eq.~(\ref{eq:rot_Dirac}) we disregard some negligible terms~\cite{Muller76}.

The wave function is represented as
\begin{equation}
\psi^k(\rho, z, \varphi, t)=\sum_{m=-M}^M \psi^k_m (\rho, z, t) \, \Phi^k_m(\varphi) \, ,
\label{eq:m_expansion}
\end{equation}
where $m$ is a half-integer quantum number, index $k$ enumerates the components of $\psi(\rho, z, \varphi, t)$ , 
$2M+1$ is the number of $m$-states under consideration,
and
\begin{equation}
  \Phi^k_m(\varphi)=\frac{1}{\sqrt{2\pi}} \left\{
   \begin{array}{lc}
 	  \exp \left[ i \left( m-1/2 \right)\varphi \right], \, k=1,3 \\ [0.8mm]
 	  \exp \left[ i \left( m+1/2 \right)\varphi \right], \, k=2,4  \,. 
    \end{array}
  \right.
 \end{equation}
 Each component $\psi^k_m(\rho, z, t)$ is expanded in the basis of cubic Hermite splines $\left\{ \phi_i(\rho,z) \right\}$
\begin{equation}
\psi^k(\rho, z, \varphi, t)=\sum_{m=-M}^M \sum_{i=1}^N C^k_{i,m} (t) \, \phi_i(\rho,z) \, \Phi^k_m(\varphi) \, .
\label{eq:basis_expansion}
\end{equation}
Let us introduce the new notation $C_j \equiv  C^k_{i,m}$, $f_j \equiv \phi_i \, \Phi^k_m$. The coefficients $C_j$
can be obtained by solving the equation
\begin{equation}
iS \, \frac{d \vec{C}(t)}{dt}=H(t) \, \vec{C}(t).
\label{eq:finite_Dirac}
\end{equation}
Here $\vec{C}$ is the vector of the coefficients $C_j$, $H_{ij}=\langle f_i \arrowvert \hat{H} \arrowvert f_j \rangle$,
$S_{ij}=\langle f_i \arrowvert  f_j \rangle$. The matrix elements $S_{ij}$ and $H_{ij}$ are calculated by numerical integration
with respect to $\rho$ and $z$, and analytical integration with respect to the $\varphi$ coordinate.
For sufficiently small time interval $\Delta t$ the formal solution of Eq.~(\ref{eq:finite_Dirac}) is
 \begin{equation}
     \vec{C}(t+\Delta t)=\exp \left(-iS^{-1}H \Delta t \right) \vec{C}(t).
     \label{eq:formal_solution}
  \end{equation}
 The exponential operator is expanded as   
 \begin{equation}
  \vec{C}(t+\Delta t)=\left[ 1-iS^{-1}H \Delta t-\frac{1}{2}\left( S^{-1}H \Delta t  \right)^2+...    \right]\vec{C}(t).
  \label{eq:exponent_expansion}
\end{equation} 
 The vector $\vec{C}(t)$ is calculated on each time step using Eq.~(\ref{eq:exponent_expansion})
 with the first four terms. 
\section{Results}
\label{sec:Results}
\begin{figure}[h]
  \centering
  \includegraphics[ width=0.8\textwidth ]{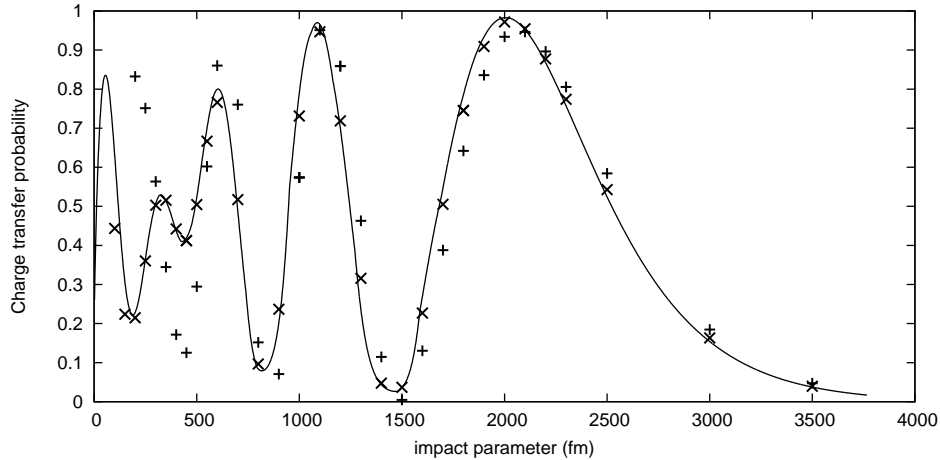}
  \caption{The charge transfer probability as a function of the impact parameter. Signs
           ``$+$'' and ``$\times$'' indicate the results of one channel and six channels calculations, respectively.
           The solid line corresponds to the results from Ref.~\cite{Tup_10}}
  \label{gr:capture}
 \end{figure}

The charge transfer probabilities in the $\rm U^{92+}$--$\rm U^{91+}(1s)$ collisions were obtained 
for one ($m=1/2$) and six channels with different $m$ (see Eq.~\ref{eq:m_expansion}) at the projectile energy $\rm E_{lab}=6$~MeV/u.
The numbers of basis functions were 15360 and 92160 for one and six channels, respectively. The results are shown in Fig.~\ref{gr:capture}.
\section{Conclusion}
In the present work the charge transfer probabilities in the $\rm U^{92+}$--$\rm U^{91+}(1s)$ collisions 
were calculated for a wide range of the impact parameter.
As can be seen from Fig.~\ref{gr:capture}, the results of the calculations with six channels are
in a good agreement with the corresponding values obtained in Ref.~\cite{Tup_10}. Since 
$\langle \Phi_m \arrowvert \vec{J} \vec{\omega} \arrowvert  \Phi_m \rangle=0$, the rotation of the internuclear axis
is not taken into account in the case of the one channel calculations. As one can observe from 
Fig.1, the influence of this rotation is essential for the impact parameter $b<$ 600 fm.  
\section{Acknowledgements}
This work was supported by RFBR (Grants No. 10-02-00450 and No. 11-
02-00943-a), by GSI, by DAAD, by the grant of the President of the Russian Federation (Grant
No. MK-2106.2012.2), and by the Russian Ministry of Education and Science 
(Program ``Scientific and scientific-pedagogical personnel of innovative Russia'', Grant No. 8420). 
YSK and IAM acknowledge financial support
from the FAIR-Russia Research Center.
The work of IAM was also supported by the
Dynasty Foundation and by G-RISC.

\end{document}